\documentclass[sigconf]{acmart}

\AtBeginDocument{%
  }

\copyrightyear{2024} 
\acmYear{2024} 
\setcopyright{rightsretained} 
\acmConference[ASE '24]{39th IEEE/ACM International Conference on Automated Software Engineering }{October 27-November 1, 2024}{Sacramento, CA, USA}
\acmBooktitle{39th IEEE/ACM International Conference on Automated Software Engineering (ASE '24), October 27-November 1, 2024, Sacramento, CA, USA}
\acmDOI{10.1145/3691620.3695350}
\acmISBN{979-8-4007-1248-7/24/10}

\author{Kristian Kolthoff}
\orcid{0000-0003-4982-488X}
\affiliation{%
  \institution{Institute for Enterprise Systems, University of Mannheim}
  \city{Mannheim}
  \country{Germany}
}
\email{kolthoff@es.uni-mannheim.de}

\author{Christian Bartelt}
\orcid{0000-0003-0426-6714}
\affiliation{%
  \institution{Institute for Enterprise Systems, University of Mannheim}
  \city{Mannheim}
  \country{Germany}
}
\email{bartelt@es.uni-mannheim.de}

\author{Simone Paolo Ponzetto}
\orcid{0000-0001-7484-2049}
\affiliation{%
  \institution{Data and Web Science Group University of Mannheim}
  \city{Mannheim}
 \country{Germany}
}
\email{simone@informatik.uni-mannheim.de}

\author{Kurt Schneider}
\orcid{0000-0002-7456-8323}
\affiliation{%
  \institution{Leibniz Universität Hannover}
  \city{Hannover}
 \country{Germany}
}
\email{kurt.schneider@inf.uni-hannover.de}


\begin{document}

\title{Self-Elicitation of Requirements with Automated GUI Prototyping}


\renewcommand{\shortauthors}{Kolthoff et al.}

\begin{abstract}
Requirements Elicitation (RE) is a crucial activity especially in the early stages of software development. GUI prototyping has widely been adopted as one of the most effective RE techniques for user-facing software systems. However, GUI prototyping requires \textit{(i)} the availability of experienced requirements analysts, \textit{(ii)} typically necessitates conducting multiple joint sessions with customers and \textit{(iii)} creates considerable manual effort. In this work, we propose \textit{SERGUI}, a novel approach enabling the Self-Elicitation of Requirements (SER) based on an automated GUI prototyping assistant. \textit{SERGUI} exploits the vast prototyping knowledge embodied in a large-scale GUI repository through Natural Language Requirements (NLR) based GUI retrieval and facilitates fast feedback through GUI prototypes. The GUI retrieval approach is closely integrated with a Large Language Model (LLM) driving the prompting-based recommendation of GUI features for the current GUI prototyping context and thus stimulating the elicitation of additional requirements. We envision \textit{SERGUI} to be employed in the initial RE phase, creating an initial GUI prototype specification to be used by the analyst as a means for communicating the requirements. To measure the effectiveness of our approach, we conducted a preliminary evaluation. Video presentation of \textit{SERGUI} at: \url{https://youtu.be/pzAAB9Uht80}

\end{abstract}



\maketitle

\section{Introduction}

Numerous techniques have been proposed and employed over the years with the aim of enhancing requirements elicitation in user-facing software systems development \cite{pohl2010requirements}. GUI prototyping serves as such an elicitation technique offering a mechanism for analysts to visualize their comprehension of the requirements and allowing customers to validate these through a tangible artifact. Additionally, these prototypes lay the groundwork for actively involving customers during the development phase. This involvement can spark meaningful discussions leading to the clarification and refinement of requirements \cite{ravid2000method, beaudouin2002prototyping}. However, conducting RE using GUI prototyping requires the availability of experienced requirements analysts and creates considerable manual efforts. Furthermore, GUI prototyping is an iterative and time-consuming process, often requiring multiple joint sessions between analysts and customers, resulting in outdated, invalid requirements \cite{schneider2007generating}.

In this work, we propose \textit{SERGUI}, a novel RE approach that enables customers to perform Self-Elicitation of Requirements (SER) for user-facing software systems through automated NL-based GUI prototyping and that provides an LLM-based GUI feature recommendation mechanism to promote the automatic elicitation of additional requirements. The general notion of SER refers to guiding customers to uncover their own requirements and has originally been introduced by \textit{LadderBot} \cite{rietz2019ladderbot}, thus reducing the effort of analysts in the initial RE phase and facilitating to closely integrate customers into the RE process. By extending the notion of SER with automated GUI prototyping, we exploit the benefits of GUI prototypes as a requirements specification artifact, facilitating to obtain fast customer feedback and achieving early clarification of requirements. In \textit{SERGUI}, customers are guided through the GUI prototyping process by a rule-based dialogue assistant which provides matching GUIs and proactively recommends potentially relevant GUI features. Our \textit{SERGUI} prototype, source code and evaluation datasets are all available at our accompanying repository \cite{sergui, sergui-github}.

\section{Approach: SERGUI}

\begin{figure*}
\includegraphics[width=\textwidth]{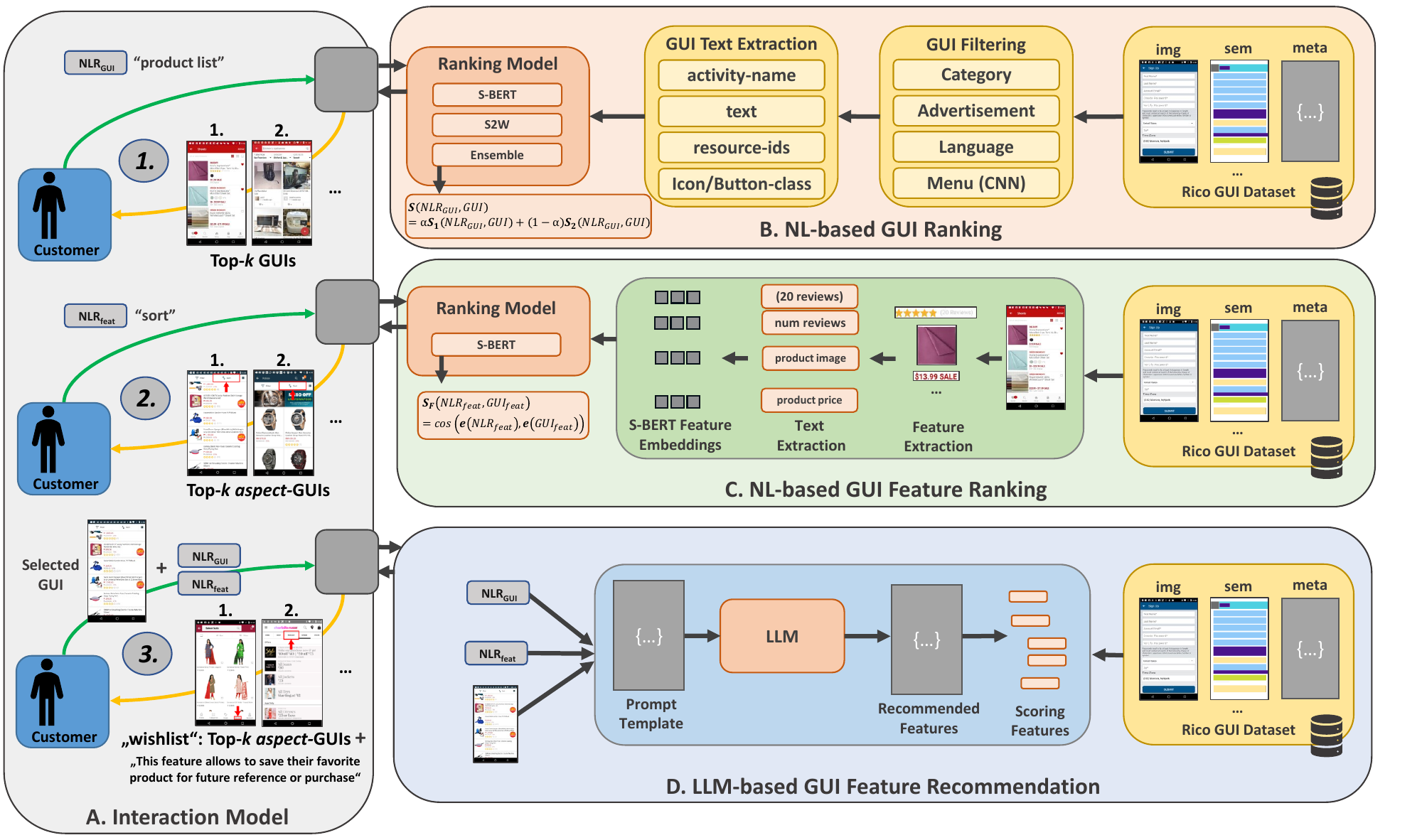}
  \caption{Overview of the \textit{SERGUI} approach with \textit{(A)} the interaction model showing the main interaction mechanisms, \textit{(B)} NL-based GUI ranking approach, \textit{(C)} the NL-based GUI feature ranking and \textit{(D)} the LLM-based GUI feature recommendation}
	\label{fig:overview}
\end{figure*}

\textit{SERGUI} is an approach to automate the initial RE phase by leveraging a comprehensive GUI repository in combination with a GUI ranking and feature recommendation technique. Subsequently, an overview of our approach is given and depicted in Fig. \ref{fig:overview}. \textit{SERGUI} is divided into multiple components: First, \textit{(A)} an interaction model encompassing the essential interaction mechanisms between the customer and the automatic GUI prototyping assistant. Second, \textit{(B)} an NL-based GUI ranking technique building upon work of \cite{kolthoff2023data, kolthoff2020gui2wire, kolthoff2021automated}. Third, \textit{(C)} an NL-based GUI feature ranking mechanism. Fourth, \textit{(D)} an LLM-based GUI feature recommendation mechanism to proactively suggest potentially relevant GUI features and simultaneously illustrate them matched on the GUIs from the top-\textit{k} GUI ranking.

\subsection{Interaction Model}

\textit{SERGUI} is built as an interactive dialogue-based approach providing guidance through the GUI prototyping process, facilitating the close integration of customers. The interaction model encompasses three essential patterns that are repeated for each GUI in the application prototype. Initially, customers can specify their NLR for a particular GUI (NLR$_{GUI}$), which is responded by the approach with a top-\textit{k} GUI ranking, as illustrated in Fig. \ref{fig:overview} \textit{A1}. This ranking represents the best matches as computed by the NL-based GUI ranking model. The GUI ranking showcases many potentially relevant GUIs encompassing potentially numerous variations, which already stimulates RE. Second, customers are enabled to specify additional NLR for individual features (NLR$_{feat}$), which is responded by the approach with a top-$k$ ranking of \textit{aspect}-GUIs, as depicted in Fig. \ref{fig:overview} \textit{A2}. Therefore, the approach presents a ranking of matching GUIs potentially containing the formulated GUI feature. If the customer finds a relevant \textit{aspect}-GUI, then it can be selected. Subsequently, the GUI ranking is recomputed based on the new aspects and the \textit{aspect}-GUI will be saved as part of the GUI prototype. Third, when the customer initially selects a GUI already satisfying many features plus the specified \textit{aspect}-GUIs, then the approach utilizes the NLR$_{GUI}$, collection of NLR$_{feat}$ and the selected GUI to proactively recommend potentially relevant GUI features, as shown in Fig. \ref{fig:overview} \textit{A3}. In an iterative fashion, the \textit{SERGUI} approach will then present each GUI feature with a short textual explanation and the top-\textit{k} ranking of potentially matching \textit{aspect}-GUIs encompassing and visualizing the respective GUI feature. Similarly to before, a relevant \textit{aspect}-GUI can directly be selected, which then will be added to the overall GUI prototype specification and exploited for GUI reranking. This GUI specification is then added to a simple linear app prototype, allowing customers to briefly skip through the app.

\subsection{NL-Based GUI Ranking}

To achieve NL-based GUI ranking, we mainly adopt the work of \cite{kolthoff2023data} and provide an extension for filtering and ranking models. The GUI ranking exploits the GUI repository \textit{Rico} \cite{deka2017rico}, the largest GUI dataset for mobile apps available. We follow the GUI filtering pipeline of \cite{kolthoff2023data}. To filter opened menus, we trained a CNN-classifier (three convolution and pooling layers) and provided as input a combination of both the original GUI screenshot image and the semantic annotation image as grayscale variants. We trained the CNN model for 6 epochs (adagrad optimizer and binary crossentropy as loss) and achieved a satisfying performance on a separate test set (\textit{Precision}=.9818 / \textit{Recall}=.7012). We focused on achieving a high precision, to avoid the removal of adequate GUIs. Overall, we filtered 23,817 GUIs (with 9,363 by the CNN) resulting in 48,402 GUIs remaining.

As a GUI ranking approach, we adopt the strong pretrained embedding-based \textit{SentenceBERT} model from \cite{kolthoff2023data}. This model computes the ranking utilizing cosine-similarity between the embedded query and embedded textual representation of the GUI i.e. score $\textbf{S$_{1}$}(NLR_{GUI}, GUI) = \cos(\textbf{e}(NLR_{GUI}), \textbf{e}(GUI))$ with $\textbf{e}$ referring to the \textit{SentenceBERT} embedding. Moreover, we extend this GUI ranking score by incorporating the \textit{S2W} dataset \cite{wang2021screen2words}, adding another artifact to the GUIs to score against. \textit{S2W} is a large collection of manually crafted high-level NL descriptions of \textit{Rico} GUIs, providing five descriptions per GUI from different annotators. For the \textit{S2W} data, we similarly compute the average over the cosine-similarity between the query and all five descriptions i.e. score $\textbf{S$_{2}$}(NLR_{GUI}, GUI) = \frac{1}{5}\sum_{i=1}^{5}\cos(\textbf{e}(NLR_{GUI}), \textbf{e}(S2W(GUI,i)))$. By creating an ensemble GUI ranking model using both the extracted GUI text and the high-level descriptions of \textit{S2W} (see formula Fig. \ref{fig:overview} \textit{B}), we facilitate obtaining a more flexible and robust GUI ranking.  

\subsection{NL-Based GUI Feature Ranking}

To achieve NL-based GUI feature ranking, we employ the top-\textit{k} GUIs representing potentially relevant context for matching the features. Here, we restrict GUI features to individual GUI components. From the top-\textit{k} GUIs, textual representations of the GUI components are extracted, including multiple texts for each component (displayed text, \textit{resource-id} and semantic classes). These text representation are then utilized to match against $NLR_{feat}$ again based on the cosine-similarity of the \textit{SentenceBERT} embeddings (see Fig. \ref{fig:overview} \textit{C}).

\subsection{LLM-Based GUI Feature Recommendation}

Within the \textit{SERGUI} approach, the GUI feature recommendations are based on few-shot prompting of an LLM (GPT-4) using the context of \textit{(i)} the initial textual requirements denoted by $NLR_{GUI}$, \textit{(ii)} a collection of already specified features by the customer as $NLR_{feat}$ and \textit{(iii)} an initially selected GUI from the top-\textit{k} GUI ranking. With these contextual inputs, we fill in a prompt template. First, \textit{(A)} we provide the \textit{task instructions} asking the model to recommend the top-\textit{30} GUI features given the described context. Next, \textit{(B)} the template displays the initial requirements as $NLR_{GUI}$. Third, \textit{(C)}, the initially selected GUI is provided to the model. Since the original XML-based GUI hierarchy contains a large amount of information, transforming the GUI hierarchy into a more abstract and focused representation is necessary. Each GUI component is transformed to a string of \textit{"uicomp-text" (uicomp-type) (resource-id)} and then arranged in a two-level list based on grouping information (prompt details in \cite{sergui-github}). To compute a GUI feature ranking, each of the predicated features is matched against the features in each GUI in the top-\textit{k} GUI ranking. To obtain a confidence score that a GUI contains a predicated feature ($NLR_{feat}$) and to determine which GUI component within the GUI matches best (as \textit{aspect}-GUI), we compute $\textbf{S$_{g}$}(NLR_{feat}, GUI) = \max_{f \in GUI_{feats}}\textbf{S$_{F}$}(NLR_{feat}, f$$)$. For each predicated GUI feature, we compute a feature score as $\textbf{S$_{pf}(NLR_{feat})$} = \frac{1}{k}\sum_{i=1}^{k}S_{g}(NLR_{feat}, GUI_{k})$ over the top-\textit{k} GUI ranking to estimate the feature coverage among top-\textit{k} GUIs. The features are then recommended in the sorted order by utilizing the score $\textbf{S$_{pf}$}$ and showing the top-\textit{k} \textit{aspect}-GUIs. Using a few-shot prompt, we generate text explanations for each predicted feature.

\begin{figure}
  \includegraphics[width=0.47\textwidth]{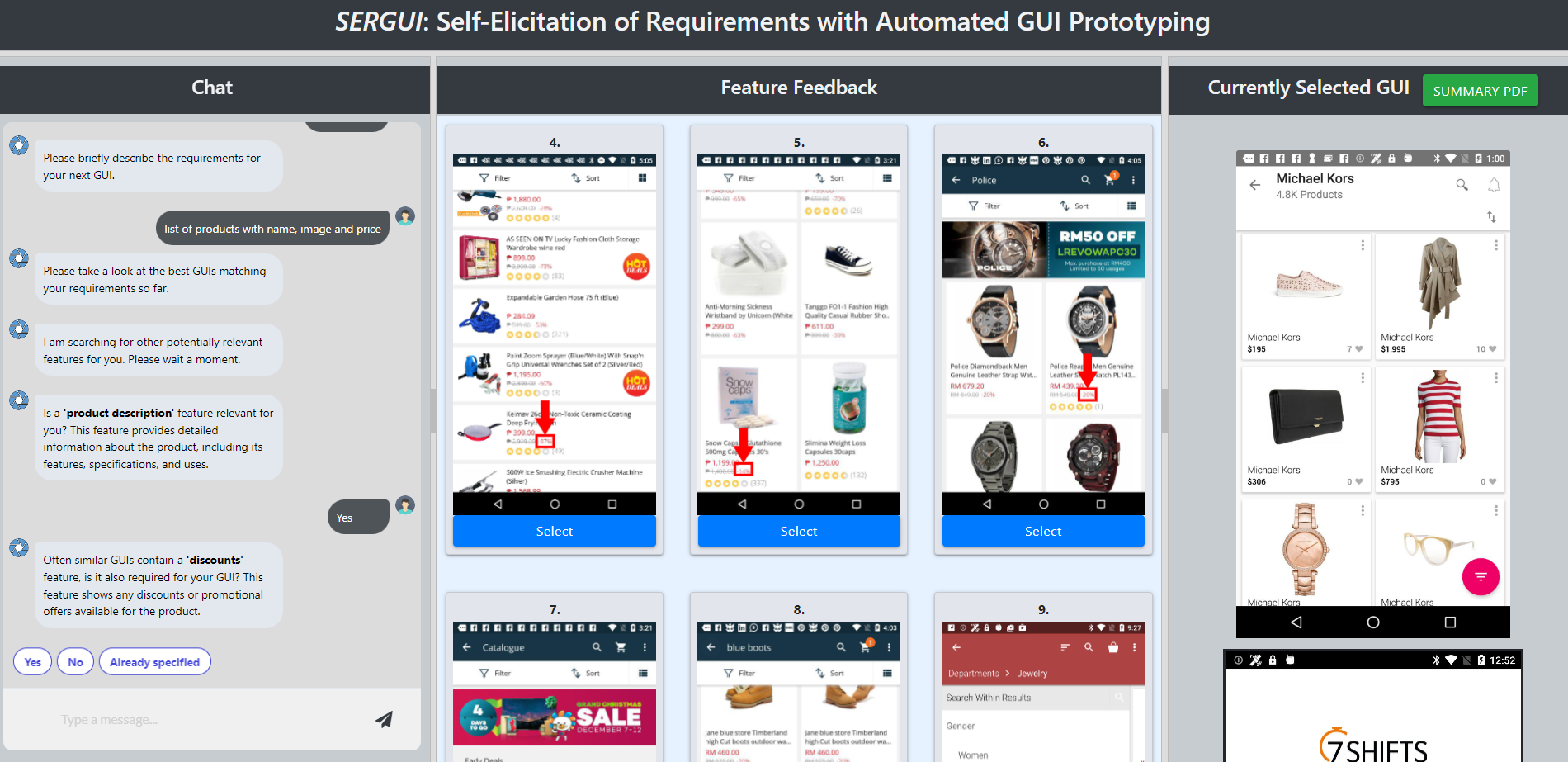}
  \caption{Web app implementation of the \textit{SERGUI} approach with a chat-section, workbench and GUI prototype preview}
	\label{fig:tool}
\end{figure}

\subsection{Feature-Based GUI Reranking}

Based on positive feedback from the customer by selecting an \textit{aspect}-GUI from the top-\textit{k} \textit{aspect}-GUI ranking, we can compute a reranking of the GUIs. Often customers may not explore much more than the top-\textit{20} or top-\textit{30} GUIs in the ranking, missing potentially relevant GUI features. By reranking the top-\textit{k} GUIs, customers potentially find more relevant GUIs at the top of the GUI ranking or they can keep their original choice. We compute the GUI reranking as $\textbf{S$_{RR}$}(GUI) = \beta \textbf{S}(NLR_{GUI}, GUI) + (1 - \beta) \frac{1}{|F|}\sum_{f \in F}{\textbf{S$_{g}$}(f, GUI)}$ ensemble of GUI ranking and normalized sum over all feature scores.

\subsection{Prototyping Artifact}

At the end of the GUI prototyping process, \textit{SERGUI} produces an app summary including \textit{(i)} a visualization of the overall app and \textit{(ii)} for each GUI the main selected GUI, the collection of \textit{aspect}-GUIs and an additional collection of textual requirements, representing recommended features that were relevant but no \textit{aspect}-GUI was found. This GUI prototyping artifact produced by \textit{SERGUI} can then be employed by analysts as a starting point for further elicitation.

\section{Experimental Evaluation}

To assess the performance of \textit{(i)} the feature recommendation, \textit{(ii)} relevance of matched \textit{aspect}-GUIs and \textit{(iii)} feature-based GUI reranking, we conducted a small preliminary evaluation. We recruited 12 annotators possessing technical backgrounds (BSc.:7|MSc.:5). In addition, the recruited participants had medium to high experience in software development (Mean:3.50|SD:0.90), as self-reported on a five-point Likert scale, allowing them to evaluate the relevance of the recommendation results. The evaluation consisted of three applications from diverse domains (\textit{shopping}, \textit{news} and \textit{social}), each encompassing three GUIs, overall resulting in nine different GUIs. Each application was presented to the participants as a low-fidelity GUI prototype, containing solely a minimal collection of features to enable participants to recognize the notion of the GUI. On this basis, participants were asked to employ \textit{SERGUI} for prototyping and evaluate the relevance of the results. Overall, 72 GUIs were prototyped during the preliminary evaluation (six per annotator).

\subsection{GUI Feature Recommendation Relevance}

To evaluate the feature recommendation relevance, 720 recommendations were made (10 per GUI) and we computed \textit{MAP}, \textit{MRR} and \textit{P@k}. A \textit{MAP} of .741 indicates a strong feature recommendation performance as approximately three out of four features are marked as relevant on average. Likewise, the \textit{MRR} of .816 describes that the first relevant feature occurs at rank 1.22 on average indicating that the top-features are relevant. In addition, also the \textit{P@k} values indicate a substantial recommendation performance starting from .691 (\textit{P@1}) to .638 (\textit{P@10}). This also shows that the feature scoring accordingly ranks more relevant documents higher as intended.

\subsection{GUI Feature Ranking Relevance}

To evaluate the relevance of the matched \textit{aspect}-GUIs, we employed the identical setup as described before (10 features per GUIs with top-\textit{15} \textit{aspect}-GUIs) and for each relevant feature (either answered by \textit{yes} or by picking an \textit{aspect}-GUI), we employed the rank of the selected \textit{aspect}-GUI to compute respective ranking metrics. Since we only have at maximum one single relevant \textit{aspect}-GUI, we computed the \textit{MRR} and \textit{HITS@k}. A \textit{MRR} of .390 shows that that the first relevant \textit{aspect}-GUI appears at rank 2.56 on average indicating a substantial performance in finding relevant \textit{aspect}-GUI matches. Moreover, starting from a \textit{HITS@1} of .270 represents that on average in 27\% a relevant \textit{aspect}-GUI could be found at the top-1 position. A \textit{HITS@15} of .691 represents that on average our matching approach could find a relevant \textit{aspect}-GUI in 69.1\% within the top-15 \textit{aspect}-GUI ranking. Therefore, for a large amount of recommended features our feature matching approach is capable of finding a relevant \textit{aspect}-GUI visualizing the predicted feature.

\subsection{GUI Reranking Performance}

By comparing the \textit{SentenceBERT} (\textit{MRR:} .172) and \textit{S2W} (\textit{MRR:} .210) ranking models with the ensemble model (\textit{MRR:} .429), we see that the ensemble considerably improves the GUI ranking performance. Based on the positive feature feedback (selecting an \textit{aspect}-GUI), we computed a reranking of the GUIs. Participants could initially select a GUI and then are presented with feature recommendations. Afterwards, participants could either chose a new GUI from the updated ranking or keep their previous choice. In 68.05\% of the cases, the participants selected a better matching GUI. The reranking score (initial - updated rank) is +61.89 ranks on average (SD:79.82|Min:-26|Max:335) indicating that the feature-based reranking can substantially improve the ranks of relevant screens.

\section{Related Work}

The \textit{Fast Feedback} technique \cite{schneider2007generating} reduces the number of necessary RE sessions with the customer by providing a tool-based support to rapidly and interactively create pen-paper prototypes combined with use cases and directly allows to incorporate customer feedback. However, still an experienced analyst is required to conduct the initial elicitation.  Recent approaches \textit{GUI2WiRe} \cite{kolthoff2020gui2wire, kolthoff2021automated} and \textit{RaWi} \cite{kolthoff2023data} support analysts to rapidly create GUI prototypes via NL-based GUI retrieval, however, cannot readily be employed by customers due to the lack of knowledge and experience with GUI prototyping and hence are dependent on analysts. In the RE approach in \cite{teixeira2014requirements}, analysts create initial prototypes based on a start-up meeting with the clients, subsequently enabling the customers to modify the prototype through a web-based tool. Similarly, the \textit{Graphical Requirements Collector} \cite{moore2000comparison} requests users to directly create their own GUI prototypes by constructing them bottom-up and augmenting them with textual requirements. This integrates customers closely into the process, however, due to the lack of guidance, requires customers to have technical knowledge and experience with GUI prototyping. Moreover, \textit{LadderBot} \cite{rietz2019ladderbot} is able to conduct dialogue-based elicitation with customers by applying laddering, a structured interview technique. However, this automatic dialogue-based elicitation approach focuses on the collection of simple textual requirements solely and therefore cannot be applied for the initial elicitation with GUI prototyping. Consequently, this technique cannot provide fast feedback to customers on the basis of GUI prototypes like \textit{SERGUI}.

\section{Conclusion and Future Work}

In this work, we proposed \textit{SERGUI} as the first approach towards enabling the Self-Elicitation of Requirements with GUIs for customers. The evaluation results show that \textit{SERGUI} is able to effectively support users during the GUI prototyping process in terms of providing relevant GUIs on the basis of NLR, recommending relevant GUI features and their visualizations. For future work, we plan to conduct a large user study to more comprehensibly evaluate our approach.

\bibliographystyle{ACM-Reference-Format}
\bibliography{ref}

\end{document}